\documentclass{PoS}

\title{Quark and gluon confinement from an effective model of Yang-Mills theory}

\ShortTitle{Quark and gluon confinement from an effective model of Yang-Mills theory}

\author{\speaker{Kei-Ichi Kondo}%
         \thanks{This work is  supported by Grant-in-Aid for Scientific Research (C) 21540256 from Japan Society for the Promotion of Science
(JSPS).
}\\
        Department of Physics, Graduate School of Science, Chiba University, Chiba 263-8522, Japan\\
        E-mail: \email{kondok@faculty.ch5iba-u.jp}}

\abstract{
We derive a gauge-invariant low-energy effective model of the SU(2) Yang-Mills theory. 
We find that the effective gluon propagator belongs to the Gribov-Stingl type, irrespective of the gauge choice. In the maximally Abelian gauge, especially, we demonstrate that the model exhibits both quark confinement and gluon confinement: the Wilson loop average has area law and the Schwinger function violates  reflection positivity. 
Moreover, we give a formula for the string tension calculable from the gluon propagator of the gauge-invariant field strength and gives a good estimate for the string tension.
 We discuss if quark confinement and gluon confinement are of  the same origin attributed to the gluon propagator in the deep infrared momentum region. 
}

\FullConference{International Workshop on QCD Green's Functions, Confinement and Phenomenology\\
		 5-9 September 2011\\
		 Trento, Italy}

\usepackage{amsmath,amssymb}
\usepackage{mathrsfs}
\usepackage{bm}

\begin{document}

\section{Introduction}

For quark confinement, the most well known and natural criterion is the Wilson criterion: the area law for the Wilson loop average, which means a linear static potential between a pair of quark and antiquark. The Wilson criterion for quark confinement is a gauge-invariant statement, which is independent from the gauge fixing chosen in quantizing the Yang-Mills theory. 
For gluon confinement, such criterion is not yet known as far as I know. 

For color confinement, there are at least two known approaches of Kugo--Ojima (KO) \cite{KO79} and Gribov--Zwanziger (GZ) \cite{GZ78}. 
In these approaches, the criteria for color confinement are attributed to the deep infrared behavior of specific Green functions.  
However, the Green functions depend on the gauge.
Therefore, color confinement has been studied so far gauge by gauge, e.g., the Lorenz gauge, Coulomb gauge, the maximally Abelian (MA) gauge, etc. 
In the Lorenz-Landau gauge, especially, the KO criterion for color confinement is reduced to the infrared behavior of the ghost propagator \cite{KO79}.

Recent investigations show that gloun and ghost propagators in the most common Landau gauge are classified into two types according to their behaviors in the infrared region:

\begin{enumerate}

\item[
$\bullet$] scaling solution (IR suppressed gluon propagator and  enhanced ghost propagator) 
\\
\indent 
Schwinger-Dyson equation (SDE) [von Smekal, Hauck \& Alkofer,1997,1998 \cite{AvS01}]...

\item[
$\bullet$] decoupling solution (IR finite gluon propagator and  tree-like ghost propagator)
\\
\indent 
SDE [Boucaud et al.,2008 \cite{Boucaud-etal08}] [Aguilar,Binosi, \& Papavassiliou,2008 \cite{ABP08}]... 
\\
\indent 
Lattice simulations [Bogolubsky et al., 2009 \cite{BIMPS09}] 
[Cucchieri \& Mendes, 2008 \cite{CM08}]
[Sternbeck, von Smekal, Leinweber, Williams, 2007 \cite{SvSLW07}]...
\end{enumerate}
The scaling solution fulfills the KO/GZ criterion, while this is not the case for the decoupling solution.
The decoupling solution is supported by recent results of numerical simulations on the lattice with very large volumes \cite{BIMPS09,CM08,SvSLW07}.  
It is still under active debate to discriminate two different types of propagators.

No one has found a gauge-independent criterion for color confinement!
Therefore, quark confinement in the Wilsonian sense cannot be derived  at present as a special case of these color confinement scenarios. 
In fact, the relationship between quark confinement and the Green functions in the infrared regime is not yet clarified.

Nevertheless, it has been shown that  both scaling and decoupling solutions exhibit quark confinement and gluon confinement: 
\begin{enumerate}
\item[
$\bullet$] quark confinement, i.e. the vanishing of the Polyakov loop average at finite temperature $T$ for $0 < T< T_c$ below the critical temperature $T_c$.
\\
functional renormalization group (FRG) [Marhauser \& Pawlowski, 2008 \cite{MP08}]
[Braun, Gies and Pawlowski, 2010 \cite{BGP10}]
[Kondo, 2010 \cite{Kondo10}]

\item[
$\bullet$] gluon confinement, i.e., violation of reflection positivity was demonstrated

for scaling solutions: 
SDE [Alkofer and von Smekal, 2001 \cite{AvS01}] 
\\
\indent
for decoupling solutions: 
SDE,FRG [Fischer, Maas and Pawlowski, 2009 \cite{FMP09} ] 
Lattice simulations  [Bowman et al., 2007] 

\end{enumerate}

\noindent
The purpose of this talk 
is to discuss how quark confinement and gluon confinement are related to the infrared behavior of the gluon propagator in the Lorentz covariant gauge based on \cite{Kondo11}.

\begin{enumerate}
\item[
(1)]
 We derive a novel low-energy effective model of $SU(2)$ Yang-Mills theory  without fixing the original  gauge symmetry.
The resulting effective gluon propagator belongs to the Gribov-Stingl type, {\it irrespective of the gauge choice}:
\begin{equation}
 \tilde D_{\rm aa}(p) =   \frac{d_0+ d_1 p^2}{ c_0+c_1 p^2+c_2 p^4} 
 . 
\nonumber
\end{equation}

\item[
(2)]
 In MA gauge, we show that the model exhibits both quark confinement and gluon confinement simultaneously in the following sense:

\noindent
$\bullet$ quark confinement:
The Wilson loop average satisfies the area law. 

\noindent
$\bullet$ gluon confinement:
A Schwinger function (Euclidean Green's functions) for the effective gluon propagator violates the reflection positivity.


\item[
(3)]
However,  for the effective gluon propagator to agree exactly with the Gribov-Stingl form $c_0 \not= 0$, one must include either  (a) a gauge-invariant nonlocal ``mass term'' or (b) a ``mass term'' that breaks nilpotency of the BRST symmetry. 
Otherwise, we have $c_0=0$.

\item[
(4)]
We argue that quark and gluon confinement can be obtained even in the absence of such a mass term.

\end{enumerate}

In the follows, we consider only the $SU(2)$ gauge group \cite{KMS06,KMS05,Kondo06} and the extension to $SU(N)$  based on \cite{KSM08}  will be given elsewhere.

\section{Reformulating the Yang-Mills theory in terms of new variables}

(Step 1)
We transform the original variables $\mathscr{A}_\mu$ to the new variables $\bm n, c_\mu $, $\mathscr{X}_\mu$: 
$$
 \text{old variables}: \mathscr A_\mu^A(x)  \Longrightarrow \text{new variables}: (\bm n^\beta(x), c_\mu(x)   ,  \mathscr X_\mu^b(x) )  
 ,
$$
according to \cite{KMS06,KMS05,Kondo06}
\begin{align}
& \bm{n}(x) = n^A(x) T_A \quad (A=1,2,3)
\nonumber\\
 &  c_\mu(x)  =  \mathscr{A}_\mu(x)  \cdot \bm{n}(x) ,
\nonumber\\
 & \mathscr{X}_\mu(x) =   i g^{-1} [ D_\mu[\mathscr{A}] \bm{n}(x) , \bm{n}(x) ] ,
\end{align}
where $\bm{n}(x)$  is the Lie-algebra $su(2)$-valued field with a unit length, i.e., $n^A(x)n^A(x)=1$.
The so-called color direction field $\bm{n}$ is obtained in advance as a functional of the original variable $\mathscr{A}_\mu$by solving the reduction condition \cite{KMS06}, e.g., 
$
 [ \bm{n}(x) , D_\mu[\mathscr{A}]D_\mu[\mathscr{A}]\bm{n}(x) ] =0 
$.

If the original Yang-Mills field $\mathscr{A}_\mu(x) =\mathscr{A}_\mu^A(x)  T_A$  is decomposed into two pieces: $
  \mathscr{A}_\mu(x) = \mathscr{V}_\mu(x) + \mathscr{X}_\mu(x) ,
$ 
then the new variable $\mathscr{V}_\mu=\mathscr{A}_\mu-\mathscr{X}_\mu$ is a Lie-algebra $su(2)$-valued fields $\mathscr{V}_\mu(x)=\mathscr{V}_\mu^A(x)T_A$ ($A=1,2,3$) given by
\begin{align}
 \mathscr{V}_\mu(x)= c_\mu(x)\bm{n}(x) +    ig^{-1} [ \bm{n}(x) , \partial_\mu \bm{n}(x) ]  .
\label{cov1b}
\end{align}
The  variables $\mathscr{V}_\mu(x)$ satisfy the properties:
\begin{enumerate}
\item[
(i)]
 $\mathscr{V}_\mu$ has the same gauge transformation as the original field $\mathscr{A}_\mu$, i.e., $\mathscr{V}_\mu(x) \rightarrow \Omega(x) \mathscr{V}_\mu(x) \Omega(x)^\dagger + ig^{-1} \Omega(x) \partial_\mu \Omega(x)^\dagger$ 
and hence its field strength $\mathscr{F}_{\mu\nu}[\mathscr{V}]:= \partial_\mu \mathscr{V}_\nu - \partial_\nu \mathscr{V}_\mu -ig[\mathscr{V}_\mu, \mathscr{V}_\nu ]$ transforms in  the adjoint way: $\mathscr{F}_{\mu\nu}[\mathscr{V}](x) \rightarrow \Omega(x) \mathscr{F}_{\mu\nu}[\mathscr{V}] \Omega(x)^\dagger$, 

\item[
(ii)]
 $\mathscr{F}_{\mu\nu}[\mathscr{V}]$ is proportional to $\bm{n}$, i.e., $\mathscr{F}_{\mu\nu}[\mathscr{V}](x):=\bm{n} (x)G_{\mu\nu}(x)$. 
\end{enumerate}
Consequently, $G_{\mu\nu}=\bm{n} \cdot \mathscr{F}_{\mu\nu}[\mathscr{V}]$ is gauge-invariant, since the field $\bm{n}$ is constructed so that it transforms as
$\bm{n}(x) \rightarrow \Omega(x)\bm{n}(x)\Omega(x)^\dagger$. 
Remarkably, $G_{\mu\nu}$ has the same form as the 't Hooft-Polyakov tensor for magnetic monopole:
\begin{equation}
 G_{\mu\nu} 
  =  \partial_\mu c_\nu - \partial_\nu c_\mu +i g^{-1} \bm{n} \cdot [\partial_\mu \bm{n} ,  \partial_\nu \bm{n} ] 
 .
 \label{G}
\end{equation}

(Step 1')
In order to obtain the dual effective theory for examining the dual superconductivity \cite{Kondo12}, we  introduce a {\it gauge-invariant} antisymmetric tensor field $({}^*B)_{\mu\nu}$ of rank 2 by inserting a unity into the path-integral \cite{Kondo97,Ellwanger98,Kondo00}:
\begin{align}
 1 =  \int \mathcal{D}B \exp \Big[ - \int d^D x
\frac{\gamma}{4} \{   ({}^*B)_{\mu\nu} 
- (\alpha \bm{n} \cdot \mathscr{F}_{\mu\nu}[\mathscr{V}]  - \beta \bm{n} \cdot i g [ \mathscr{X}_\mu , \mathscr{X}_\nu ]) \}^2  \Big] 
 ,
\end{align}
where $*$ is the Hodge dual operation. 
Here  (too many) parameters $\gamma, \alpha, \beta$ are introduced to see effects of each term. 
Putting $\gamma=0$ is a simple way of reproducing the original theory without the antisymmetric tensor field $B$.
When $\beta=\gamma^{-1}=\tilde G$ and $\alpha=0$, indeed, $({}^*B)_{\mu\nu}$ is regarded as a collective field for the composite operator $\bm{n} \cdot i g [ \mathscr{X}_\mu , \mathscr{X}_\nu ]$ with the propagator $\tilde G$ obtainable in a self-consistent way  \cite{EW94} according to the Wilsonian renormalization group (RG) \cite{Wetterich93}.  

Then the Euclidean Yang-Mills Lagrangian $\mathscr{L}_{\rm YM}[\mathscr{A} ]
=  \frac{1}{4}  (\mathscr{F}_{\mu\nu}^A[\mathscr{A}])^2$ is rewritten and modified into   
\begin{align}
 & \mathscr{L}_{\rm YM}[\mathscr{V},\mathscr{X},B]
\nonumber\\&
=      \frac{1+\gamma \alpha^2}{4}  G_{\mu\nu}^2
+ \frac{\gamma}{4}  ({}^*B)_{\mu\nu}^2 - \frac{\gamma \alpha}{2}  ({}^*B)_{\mu\nu}   G_{\mu\nu} 
+  \frac{1}{2} \mathscr{X}^{\mu A}  Q_{\mu\nu}^{AB} \mathscr{X}^{\nu B} 
+   \frac{1+\gamma\beta^2}{4} (i g [ \mathscr{X}_\mu , \mathscr{X}_\nu ])^2
 ,
\end{align}
where we have defined 
\begin{align}
Q_{\mu\nu}^{AB} :=& S^{AB} \delta_{\mu\nu} 
+ (2+\gamma\alpha\beta)g\epsilon^{ABC} n^C  G_{\mu\nu} 
- \gamma\beta g\epsilon^{ABC} n^{C}  (*B)_{\mu\nu} 
 ,
 \nonumber\\
 S^{AB} :=& - (D_\rho[\mathscr{V}]D_\rho[\mathscr{V}])^{AB}
  ,
\end{align}
with the covariant derivative $D_\mu$ in the adjoint representation with $\mathscr{V}_\mu:=\mathscr{V}_\mu^C T_C$, $(T_C)^{AB}=if^{ACB}$:
$
    D_\mu^{AB} := \partial_\mu \delta^{AB} -g f^{ABC} \mathscr{V}_\mu^C 
= [\partial_\mu  \mathbf{1} -ig \mathscr{V}_\mu]^{AB}  
$.

\section{Deriving an effective model by eliminating high-energy modes }

(Step2')
We identify $\mathscr{X}_\mu$ with the ``high-energy'' mode in the range $p^2 \in [M^2, \Lambda^2]$ and proceed to integrate out the ``high-energy'' modes $\mathscr{X}_\mu$.
Here $M$ is the infrared (IR) cutoff and $\Lambda$ is the ultraviolet (UV) cutoff as the initial value for the Wilsonian RG.

In the derivation of our effective model, we neglect quartic self-interactions among $\mathscr{X}_\mu$, i.e., $(i g [ \mathscr{X}_\mu , \mathscr{X}_\nu ])^2$.
However, we can take into account an effect coming from the quartic interaction, which influences our effective model. 
In fact, it is shown \cite{Kondo06,KKMSS05} that the quartic gluon interaction $(i g [ \mathscr{X}_\mu , \mathscr{X}_\nu ])^2$ among  $\mathscr{X}_\mu$ gluons can induce a contribution to the mass term  
\begin{equation}
 \frac12 M^2 \mathscr{X}_\mu \mathscr{X}_\mu
 \label{Mass}
\end{equation}  
through a gauge-invariant vacuum condensation  of ``mass dimension-2"  (the BRST-invariant version was proposed in \cite{Kondo01}), 
\begin{equation}
 \left< \mathscr{X}_\nu^B(x) \mathscr{X}_\nu^B(x) \right> \not= 0 ,
\end{equation}
which leads to the mass term (\ref{Mass}) with $M^2 \simeq \left< \mathscr{X}_\nu^B(x) \mathscr{X}_\nu^B(x) \right>$ up to a numerical factor.
This result is easily understood by a Hartree-Fock argument. 
This effect is included in the heat kernel calculation through the infrared regularization \cite{Kondo11,Kondo12}.

The correlation functions for new variables have been computed on a lattice by numerical simulations using the Monte-Carlo method in \cite{SKKMSI07} based on \cite{KKMSSI05,IKKMSS06}. 
This justifies the identification of $\mathscr{X}_\mu$ as the high-energy mode  negligible in the low-energy regime below $M \simeq 1.2{\rm GeV}$.  
Here the Landau gauge $\partial^\mu \mathscr{A}_\mu=0$ was adopted, since we need to fix the gauge to obtain the propagator or correlation functions. 
\footnote{
In the MA gauge, it has been shown in an analytical way \cite{DGLSSSV04} that the off-diagonal gluon mass generation can follow from the   off-diagonal gluon-ghost condensation of mass dimension 2, 
$\left< g^2 \mathcal{O} \right> ,
 \mathcal{O} :=  \frac12 A_\mu^a A^{\mu a} + \alpha \bar{C}^a C^a $ which has been proposed in \cite{Kondo01}.
} 
 
In these approximations, we can integrate out $\mathscr{X}_\mu$ by the Gaussian integration and obtain a \textit{gauge-invariant} low-energy effective action $S_{\rm YM}^{\rm eff}[\mathscr{V},B]$ 
\begin{align}
   S_{\rm YM}^{\rm eff}[\mathscr{V},B]
=  \int  \Big[  \frac{1+\gamma \alpha^2}{4}  G_{\rho\sigma}^2
+ \frac{\gamma}{4} ({}^*B)_{\rho\sigma}^2 
- \frac{\gamma \alpha}{2}  ({}^*B)_{\rho\sigma}   G_{\rho\sigma} \Big]
+  \frac{1}{2} \ln \det  Q_{\rho\sigma}^{AB}  
- \ln \det  S^{AB}
 ,
 \label{Seff}
\end{align}
where $\int =\int d^4x $, the functional logarithmic determinant $\frac{1}{2} \ln \det  Q_{\rho\sigma}^{AB}$ comes from   integrating out the $\mathscr{X}$ field, and the last term comes from  the FP-like determinant term \cite{KMS05}  associated with the reduction condition \cite{KMS06}.
We can obtain (see a subsequent paper \cite{Kondo12} for details of calculations)   
\begin{align}
  \frac{1}{2} \ln \det  Q_{\rho\sigma}^{AB}  
- \ln \det  S^{AB}
&=  \int   \frac{g^2\ln  \frac{\mu^2}{M^2}}{(4\pi )^{2}}  \left[
\frac{1}{6} G_{\rho\sigma}^2 -  \frac12 \{ (2+\gamma \alpha\beta) G_{\rho\sigma}-\gamma\beta ({}^*B)_{\rho\sigma} \}^2  
\right]
\nonumber\\&
+  \int    \frac{g^2}{(4\pi )^{2}}  \frac{1}{M^2} \frac{1}{6}   (\partial_\lambda     \{(2+\gamma \alpha\beta) G_{\rho\sigma}-\gamma\beta ({}^*B)_{\rho\sigma} \}     )^2
+ O(\partial^4/M^4)
 .
 \label{lndet}
\end{align}
The gauge fixing is unnecessary in this calculation. Indeed, the resulting effective action (\ref{Seff}) with (\ref{lndet})  is manifestly gauge-invariant.
This is one of main results.  
The correct RG $\beta$-function at the one-loop level 
$\beta(g) :=\mu \frac{dg(\mu)}{d\mu}=-b_1g^3+O(g^5)$, $b_1=\frac{22}{3}/(4\pi)^2$
is reproduced in a gauge-invariant way when $\gamma\alpha\beta=0$  which follows from e.g. $\alpha=0$ (a choice mentioned above) or $\gamma=0$ (in the case of no $B_{\mu\nu}$ field).

Thus we obtain the following effective action $S_{\rm YM}^{\rm eff}[G,B]$ up to terms quadratic in the fields, 
\begin{align}
  S_{\rm YM}^{\rm eff}[G,B] 
=& \frac{1}{2} (G,\Big[  f_0 + f_1 \varDelta \Big] G)    
   +  \frac{1}{2} ({}^*B,\left[ d_0 + d_1 \varDelta \right] {}^*B) 
   +     ( G, \Big[ h_0 + h_1  \varDelta   \Big] {}^*B) 
    + O(\frac{1}{M^4}) 
 ,
 \label{eff-action-1}
\end{align}
where
\begin{align}
d_0 =& \gamma - \frac{g^2 \ln \frac{\mu^2}{M^2}}{(4\pi )^{2}} 2\gamma^2\beta^2
 ,
\quad
d_1 =    \frac{g^2}{(4\pi )^{2}}  \frac{1}{M^2}  \frac{\gamma^2\beta^2}{3} 
 ,
\nonumber\\
f_0 =& 1+\gamma\alpha^2
-  \frac{g^2 \ln \frac{\mu^2}{M^2}}{(4\pi )^{2}} \frac{2-6(2+\gamma\alpha\beta)^2}{3}
 ,
\quad
f_1 =   \frac{g^2}{(4\pi )^{2}}  \frac{1}{M^2} \frac{(2+\gamma\alpha\beta)^2}{3} ,
\nonumber\\
h_0 =& -\gamma\alpha + \frac{g^2 \ln \frac{\mu^2}{M^2}}{(4\pi )^{2}} 2(2+\gamma\alpha\beta)\gamma\beta 
 ,
\quad
h_1 =   - \frac{g^2}{(4\pi )^{2}}  \frac{1}{M^2} \frac{(2+\gamma\alpha\beta)\gamma\beta}{3}
  .
\end{align}

We will see that the exact Gribov-Stingl form of the gluon propagator is obtained, 
if one introduces a gauge-invariant, but nonlocal ``mass term'':  
\begin{align}
& S^{\rm mass}_{\rm YM}[G] 
= \frac{1}{2}  \left(  G,  m^2\varDelta^{-1}G \right) ,
\label{mass1}
\end{align}
or, if one introduces a non gauge-invariant mass term 
\begin{align}
& S^{\rm mass}_{\rm YM}[G] 
 = \frac{1}{2}  \left(  a,  m^2 a \right) .
\label{mass2}
\end{align}
for the gauge field $a$ related to the field strength $G$ by 
\begin{equation}
 G=da , \quad ( \delta a=0 ). 
\end{equation}
\noindent
Even after taking  specific gauges, the BRST invariance is  also broken by including this mass term.
However, we can modify the BRST  such that the modified BRST is a symmetry of the Yang-Mills theory with the mass term at the cost of nilpotency. 
In other words, the requirement of nilpotency of the BRST  excludes such a gluon mass term.

\section{The Gribov-Stingl form for gluon propagator}

We can integrate out $B$ by the Gaussian integration.
Then we obtain the effective action in terms of   $G$: 
\begin{equation}
S^{\rm eff}_{\rm YM}[G] 
= \frac{1}{2} \left(  G, \mathscr{D}_{\rm GG}^{-1} G \right) 
 \label{eff-action-G}
\end{equation}
If we include the mass term (\ref{mass1}), the inverse effective propagators for the field strength $G$ reads
\begin{align}
\mathscr{D}_{\rm GG}^{-1} =&[m^2 \varDelta^{-1} + f_0 + f_1 \varDelta]-[d_0+d_1\varDelta]^{-1}[h_0+h_1\varDelta]^2 
\nonumber\\
=& \frac{[d_0+d_1\varDelta][m^2 \varDelta^{-1} + f_0 + f_1 \varDelta]-[h_0+h_1\varDelta]^2}{d_0+d_1\varDelta}  .
\end{align}
Then we obtain the effective propagator $\mathscr{D}_{  aa}^{-1}$ for the field $a$ defined by $G=da$:
\begin{equation}
 \mathscr{D}_{\rm GG}^{-1}  = \Delta^{-1} \mathscr{D}_{  aa}^{-1}  ,
 \quad
 \mathscr{D}_{ aa}^{-1}
 =  \frac{ c_0+c_1 \Delta+c_2 \Delta^2}{d_0+ d_1 \Delta} 
 ,
  \label{LEET-c-propam}
\end{equation}
where
\begin{align}
c_0 =&  m^2 d_0
 , \quad
c_1 =  d_0 f_0 -h_0^2 + m^2 d_1
 , \quad
c_2 =  d_0 f_1 + f_0 d_1  -2h_0 h_1 
 .
\end{align}
We observe that the effective gluon propagator $\mathscr{D}_{ aa}$ has the Gribov-Stingl form when $c_0 \not=0$:
\begin{equation}
 \tilde{\mathscr{D}}_{ aa}(p) =   \frac{1+ d_1 p^2}{ c_0+c_1 p^2+c_2 p^4} 
 .
  \label{LEET-a-propa}
\end{equation}
Thus, we have found that  the effective propagator has the Gribov-Stingl form \cite{Stingl86}.
Note that $d_1$ comes from the induced kinetic term for the $B$ field which was introduced in the beginning as an auxiliary field without the kinetic term. 

\section{ Converting the Wilson loop to the surface-integral}

(Step 3)
We use a non-Abelian Stokes theorem \cite{DP89,Kondo98b,Kondo08}  to rewrite a non-Abelian Wilson loop operator 
\begin{align}
  W_C[\mathscr{A}]  
:=& {\rm tr} \left[ \mathscr{P} \exp \left\{ ig  \oint_{C} dx^\mu \mathscr{A}_\mu(x) \right\} \right]/{\rm tr}({\bf 1})  
 ,
\end{align}
into the area-integral over the surface $\Sigma$ ($\partial \Sigma=C$):
\begin{equation}
 W_C[\mathscr{A}] = \int d\mu_{\Sigma}(\xi) \exp \left[  ig \frac12 \int_{\Sigma: \partial \Sigma=C} G \right] ,
 \quad
 d\mu_{\Sigma}(\xi):=\prod_{x \in \Sigma} d\mu(\xi_{x}) ,
 \label{W[V]}
\end{equation}
where $d\mu$ is an invariant measure  on $SU(2)$  normalized as $\int d\mu(\xi_{x})=1$, $\xi_{x} \in SU(2)$. 
In the two-form $G:=\frac12 G_{\mu\nu}(x) dx^\mu \wedge dx^\nu$, $G_{\mu\nu}$ agrees with the field strength  (\ref{G}) under the identification of the color field $\bm{n}(x)$ with a normalized traceless field (See also \cite{KSSK11})    
\begin{equation}
 \bm{n}(x) :=  \xi_{x} (\sigma_3/2) \xi_{x}^\dagger .
\end{equation}

Using the vorticity tensor  $\Theta_{\Sigma}$ with the support on the surface $\Sigma$ whose boundary is the loop $C$:
\begin{align}
 \Theta^{\mu\nu}_\Sigma(x) = \int_{\Sigma} d^2S^{\mu\nu}(x(\sigma)) \delta^D(x-x(\sigma)) ,
\end{align}
the surface integral  is cast into the volume integral and the Wilson loop operator is rewritten as  
\begin{equation}
 W_C[\mathscr{A}] = \int d\mu_{\Sigma}(\xi) \exp \left[  ig \frac12 (\Theta_{\Sigma},G)  \right] 
, \quad
(\Theta_{\Sigma},G)  =   \int d^Dx \frac12 \Theta_\Sigma^{\mu\nu}(x) G_{\mu\nu}(x)
 ,
 \label{W[V]2}
\end{equation}
where $( \cdot , \cdot )$ is the $L^2$ inner product for two differential forms.

\section{ Calculating the Wilson loop average to show  area law: quark confinement }

(Step 4)
We proceed to evaluate the Wilson loop average $W(C)=\langle W_{C}[\mathscr{A}] \rangle_{\rm YM}$ by using the effective action $S_{\rm YM}^{\rm eff}[G,B]$, i.e., $\langle W_{C}[\mathscr{A}] \rangle_{\rm YM}
\simeq  \langle W_C[\mathscr{A}] \rangle_{\rm YM}^{\rm eff} $ with the aid of (\ref{W[V]2}).

We demonstrate that the simplest way to obtain the area law is to use the low-energy effective action $S^{\rm eff}_{\rm YM}[G,B]$ retained up to terms quadratic and bilinear in  $G$ and $B$.

In what follows, we take the unitary-like gauge 
\begin{equation}
 n^A(x) = \delta_{A3} ,
 \label{unitary}
\end{equation}
which reproduces the same effect as taking the MA gauge \cite{tHooft81} in the original Yang-Mills theory.  In this gauge, $\mathscr{X}_\mu^A(x)$ reduces to the off-diagonal component $A_\mu^a(x)$ ($a=1,2$), while $\mathscr{V}_\mu^A(x)$ reduces to the diagonal one $A_\mu^3(x)=a_\mu(x)$, i.e.,
$\mathscr{X}_\mu^A(x)=\mathscr{A}_\mu^a(x) \delta_{Aa} $, $\mathscr{V}_\mu^A(x)=\mathscr{A}_\mu^3(x) \delta_{A3} =c_\mu(x) \delta_{A3} $. 
The gauge (\ref{unitary}) forces  the color field at each spacetime point to take the  same direction by gauge rotations. Hence the field $G$ given by (\ref{G}),
\begin{equation}
 G_{\mu\nu}  = F_{\mu\nu}+H_{\mu\nu} , \quad 
F_{\mu\nu}   :=  \partial_\mu c_\nu  - \partial_\nu c_\mu , 
\quad
H_{\mu\nu} :=  i g^{-1} \bm{n} \cdot [\partial_\mu \bm{n} ,  \partial_\nu \bm{n} ]
\end{equation}
contains singularities (of hedgehog type) similar to the Dirac magnetic monopole after taking the gauge (\ref{unitary}).
If we do not fix the gauge, such a contribution is contained also in  the part $i g^{-1} \bm{n} \cdot [\partial_\mu \bm{n} ,  \partial_\nu \bm{n} ]$ to make a gauge-invariant  combination $G_{\mu\nu}$, see \cite{KKMSSI05,IKKMSS06}.
Consequently, the Bianchi identity for $G$ is violated, 
\begin{equation}
\delta *G = *dG = **ddc+*dH = \delta * H \ne 0 ,
\end{equation}
even if $ddc=0$. 
Here $d$ denotes the exterior differential and $\delta$ the codifferential. 
There is no well-defined one-form $h$ such that $H=dh$.
Thus we obtain a nontrivial gauge-invariant magnetic monopole current defined by
\begin{equation}
 k := \delta *G .
\end{equation}

By integrating out the $B$ field, we obtain the effective action $\tilde S^{\rm eff}_{\rm YM}[G]$.
Then we find that the effective propagator $\mathscr{D}_{\rm aa}$ has the Gribov-Stingl form: 
\begin{equation}
 \tilde{\mathscr{D}}_{\rm GG}(p) = p^2 \tilde{\mathscr{D}}_{\rm aa}(p) , \
 \tilde{\mathscr{D}}_{\rm aa}(p) = \frac{1+ d_1 p^2}{ c_0+c_1 p^2+c_2 p^4} 
 ,
  \label{LEET-c-propa2}
\end{equation}
where
$c_0=m^2$,
$c_1=1+ \frac{\gamma\beta^2}{3}  \frac{g^2 }{(4\pi )^{2}}  \frac{m^2}{M^2}$,
$c_2=\frac{g^2}{(4\pi )^{2}}  \frac{1}{M^2}  [(2+\gamma\alpha\beta)^2+(1+\gamma\alpha^2)\gamma\beta^2+2(2+\gamma\alpha\beta)\gamma\alpha\beta]/3$,
and
$d_1= \frac{\gamma\beta^2}{3}  \frac{g^2}{(4\pi )^{2}}  \frac{1}{M^2}$.
The precise values of the parameters $m, \gamma, \alpha, \beta$ and $M$ are to be determined by the functional RG \cite{Wetterich93} following \cite{Kondo10}, which is a subject of   future study.

In the unitary-like gauge (\ref{unitary}) the Wilson loop operator is reduced to
\begin{equation}
    W_C[F] 
=   \exp \left[  ig \frac12 \int_{\Sigma: \partial \Sigma=C} G \right]
= \exp \left[  ig  \frac12(\Theta_{\Sigma},G)  \right] .
\end{equation}

Then the Wilson loop average $W(C)$ is evaluated by integrating out $G=da$:
\begin{align}
 W(C) = \exp \left[
-\frac18 g^2 ( \Theta_{\Sigma}, \mathscr{D}_{\rm GG}  \Theta_{\Sigma}) + ....
 \right] ,
\end{align}
where $\mathscr{D}_{\rm GG}=\varDelta\mathscr{D}_{\rm aa}$ and its Fourier transform obeys $\tilde{\mathscr{D}}_{\rm GG}(p)=p^2\tilde{\mathscr{D}}_{\rm aa}(p)$.

For concreteness, we choose $\Theta_{\Sigma}$ for a planar surface bounded by a rectangular loop $C$ with side lengths $T$ and $R$ in the $x_3-x_4$ plane.  Then we find that the Wilson loop average has the area law for  large $R$ and $T$:
\begin{equation}
W(C) \sim \exp [-\sigma RT] ,
\end{equation} 
with the string tension given by the formula:
\begin{equation}
  \sigma = \frac18 g^2 \int_{p_1^2+p_2^2 \le M^2} {dp_1dp_2 \over (2\pi)^2}   
\tilde{\mathscr{D}}_{\rm GG}(p_1,p_2,0,0) >0
  ,
  \label{st}
\end{equation}
where the momentum integration is restricted to the two-dimensional momentum space (the dimensional reduction by two \cite{Kondo98a}) and is cutoff at $M$ which is the upper limit of the low-energy effective model being meaningful.
A positive and finite string tension $0< \sigma < \infty$ 
 follows from the condition of no real poles in the effective gluon propagator $\tilde{\mathscr{D}}_{\rm GG}(p)$ in the Euclidean region, $0< \tilde{\mathscr{D}}_{\rm GG}(p)=p^2 \tilde{\mathscr{D}}_{\rm aa}(p) < \infty$, which is connected to the gluon confinement shown below.
This is another  main result.

\begin{figure}[t]
\begin{center}
\includegraphics[height=3.0cm,width=6.0cm]{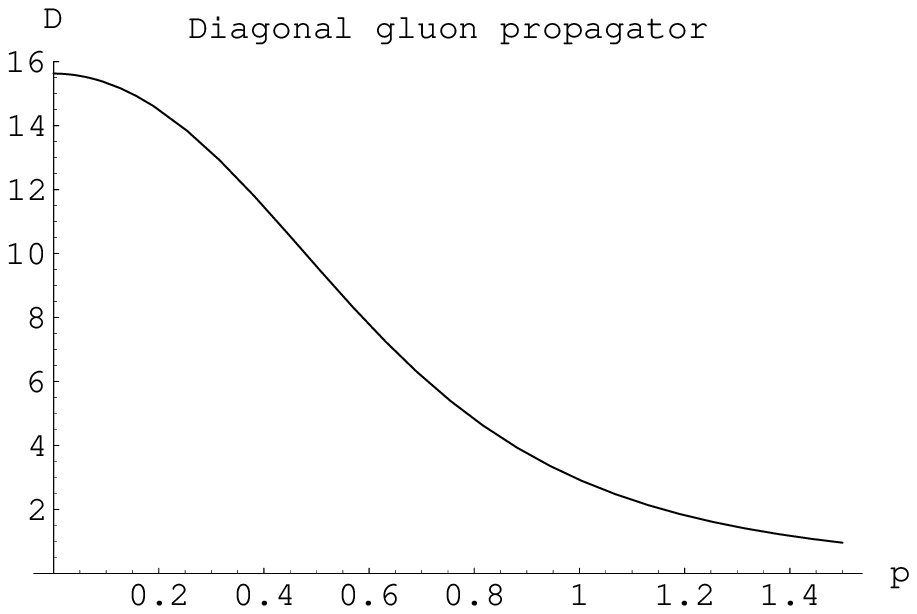}
\includegraphics[height=3.0cm,width=6.0cm]{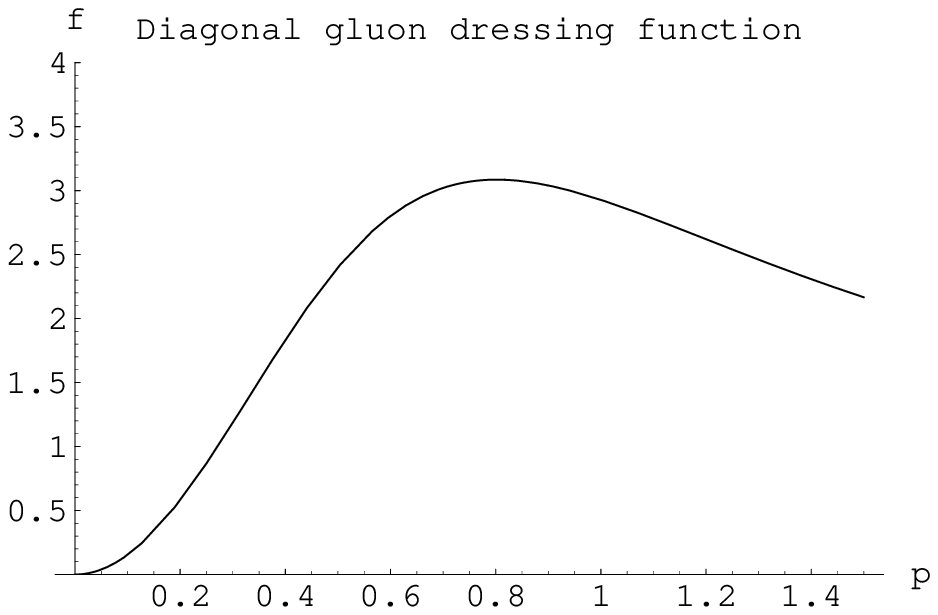}
\end{center}
\caption{
(Left panel)
The propagator of the diagonal gluon in MA gauge as a function of the momentum $p=\sqrt{p^2}$.
(Right panel)
The dressing function of the diagonal gluon in MA gauge, or  
the integrand of the formula for the string tension,
$
\frac{|p|^2+ d_1 |p|^4}{ c_0+c_1 |p|^2+c_2 |p|^4}
$
as a function of  $p=\sqrt{p^2}$.
}
\label{fig:int-st}
\end{figure}

According to numerical simulations in MA gauge \cite{AS99,BCGMP03,MCM06}, the diagonal gluon propagator is well fitted to the form (\ref{LEET-c-propa2}), see Fig.~1: e.g. \cite{MCM06} gives
$c_0= 0.064(2){\rm GeV}^2$,
$c_1= 0.125(9)$,
$c_2= 0.197(9){\rm GeV}^{-2}$,
$d_1= 0.13(1){\rm GeV}^{-2}$,
and 
$M \simeq 0.97{\rm GeV}$,
where $M$ is the mass of  off-diagonal gluons  obtained  in the MA gauge. This value of $M$ is a little bit smaller than the values of other groups \cite{AS99,BCGMP03}.
This indeed leads to a good estimate for the string tension  
\begin{equation}
 \sigma \simeq (0.5{\rm GeV})^2 ,
\end{equation} 
 according to  (\ref{st}) for    $\alpha(\mu)=g^2(\mu)/(4\pi) \simeq 1.0$ at $\mu=M$. 
The next task is to study how the results are sensitive to  the deep infrared behavior of the diagonal gluon propagator (\ref{LEET-c-propa2})  and the actual value of $M$ for the off-diagonal gluon propagator.

The Gribov-Stingl form is obtained only when $c_0 \ne 0$ (i.e., $m \ne 0$) and $d_1 \ne 0$ ($B_{\mu\nu}$ is included).
Even in the limit $m^2  \rightarrow 0$ $(c_0 \rightarrow 0)$, the area law can survive  according to (\ref{st}), provided that  
$\tilde{\mathscr{D}}_{\rm GG}(p)$ remains positive and finite: 
$ 
 \tilde{\mathscr{D}}_{\rm GG}(p) 
\rightarrow 
    \frac{ 1+ d_1 p^2}{  c_1 +c_2 p^2} 
$, 
while  $\tilde{\mathscr{D}}_{\rm aa}(p)$ behaves unexpectedly as 
$
 \tilde{\mathscr{D}}_{\rm aa}(p) \rightarrow    \frac{1+ d_1 p^2}{p^2(c_1+c_2p^2)}
$.
Hence, we  argue that it does not matter to quark confinement whether $m=0$ or $m \ne 0$.

\section{ Calculating the Schwinger function to show positivity violation: gluon confinement }

(Step5)
The positivity violation is examined. 
We consider the Schwinger function defined by
\begin{equation}
\Delta(t) 
:= \int d^3x   e^{-i\bm{p} \cdot \bm{x}}  \mathscr{D}(t,\bm{x})|_{\bm{p}=0}
= \int_{-\infty}^{+\infty} \frac{dp_4}{2\pi}  e^{ip_4 t} \tilde{\mathscr{D}} (\bm{p}=0,p_4) .
\end{equation} 
The Euclidean propagator $\tilde{\mathscr{D}}(p)$ in momentum space has a spectral representation, 
\begin{equation}
\tilde{\mathscr{D}}(p) = \int_{0}^{\infty} d\kappa^2 \frac{\rho(\kappa^2)}{p^2+\kappa^2} >0
\Longrightarrow
\Delta(t)  =  \int_{0}^{\infty} d\kappa   \rho(\kappa^2) e^{-\kappa t} > 0 .
\end{equation}
If $\Delta(t)$ is found to be non-positive, $\rho(\kappa^2)$ cannot be a positive spectral function. The corresponding states cannot appear in the physical particle spectrum: they are confined.
\\
\noindent
1) For the  free massive propagator, $\Delta(t)$ is positive for any $t$, 
\begin{align}
  \tilde{\mathscr{D}}(p) = \frac{1}{p^2+m^2}  
\Longrightarrow
  \Delta(t) 
=  \int_{-\infty}^{+\infty} \frac{dp_4}{2\pi}  e^{ip_4 t} \frac{1}{p_4^2+m^2} 
= \frac{1}{2m} e^{-m|t|} > 0 
 .
\end{align}
Therefore, we find no positivity violation as expected. 
This case corresponds to $\rho(\kappa^2)=\delta(\kappa^2-m^2)=\frac{1}{2m}\delta(\kappa-m) > 0$.

2) We consider the propagator of the Gribov-Stingl  type in Euclidean space, 
\begin{equation}
\tilde{\mathscr{D}}(p) =   \frac{d_0+ d_1 p^2}{ c_0+c_1 p^2+c_2 p^4} ,
\quad p^2 \ge 0, 
\quad c_0, c_1, c_2, d_0, d_1 \in \mathbb{R} 
 .
  \label{LEET-c-propa3}
\end{equation}In the case of $c_2 = 0$, there is no positivity violation, as far as $c_0/c_1>0$. 
In the case of $c_2 \ne 0$, $\tilde{\mathscr{D}}_{\rm aa}(p)$ has a pair of complex conjugate poles at $p^2=z$ and $p^2=z^*$,  
$
  z := x +i y
$,
$
 x := - c_1/(2c_2)
$,
$
 y := \sqrt{c_0/c_2 - \left( c_1/(2c_2) \right)^2}
$. 
We find that the Schwinger function  
$
\Delta(t) 
  := \int_{-\infty}^{+\infty} \frac{dp_4}{2\pi}  e^{ip_4 t} \tilde{\mathscr{D}}_{\rm aa} (\bm{p}=0,p_4)
$
is oscillatory in $t$ and 
is negative over finite intervals in the Euclidean time $t >0$ (See Fig.~2):
\begin{align}
  \Delta(t) 
 =   \frac{1}{2c_2 |z|^{3/2}\sin (2\varphi)} e^{-t |z|^{1/2} \sin \varphi } [ \cos (t|z|^{1/2} \cos \varphi -\varphi ) 
   + d_1 |z| \cos ( t|z|^{1/2} \cos \varphi +\varphi )  ] 
 ,
\end{align}
where
$ 
z = |z|e^{2i\varphi }
$
with
$
|z|  = \left( c_0/c_2  \right)^{1/2}
$,
$
  \cos (2\varphi) 
= - \sqrt{ c_1^2/(4c_0c_2)} 
$,
and
$
  \sin (2\varphi) 
= \sqrt{1- c_1^2/(4c_0c_2)}
$.
 Therefore, the reflection positivity is violated for the gluon propagator (\ref{LEET-c-propa2}), as long as 
\begin{equation}
 0< \frac{c_1^2}{4c_0 c_2} <1 
 , 
\end{equation}
irrespective of   $d_1$. 
When $c_0 = 0$ (or $m=0$),  
\begin{equation}
  \Delta(t) 
=  -   \frac{t}{2c_1}     -  \frac{1}{2c_1} \sqrt{\frac{c_2}{c_1} } \left( 1-\frac{c_1}{c_2} d_1 \right) e^{-t \sqrt{\frac{c_1}{c_2}}} 
 .
\end{equation}
Hence, the special case $c_0=0$ also violates the positivity, if $c_1 >0$ and $c_2 >0$.
Thus the diagonal gluon in  the MA gauge can be confined. 

\begin{figure}[t]
\begin{center}
\includegraphics[height=3.0cm,width=4.0cm]{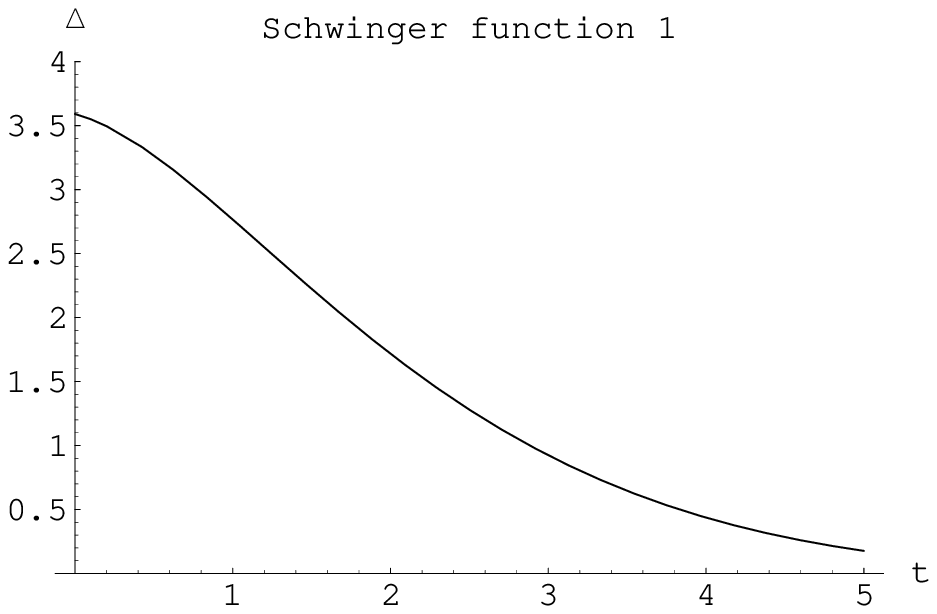}
\includegraphics[height=3.0cm,width=4.0cm]{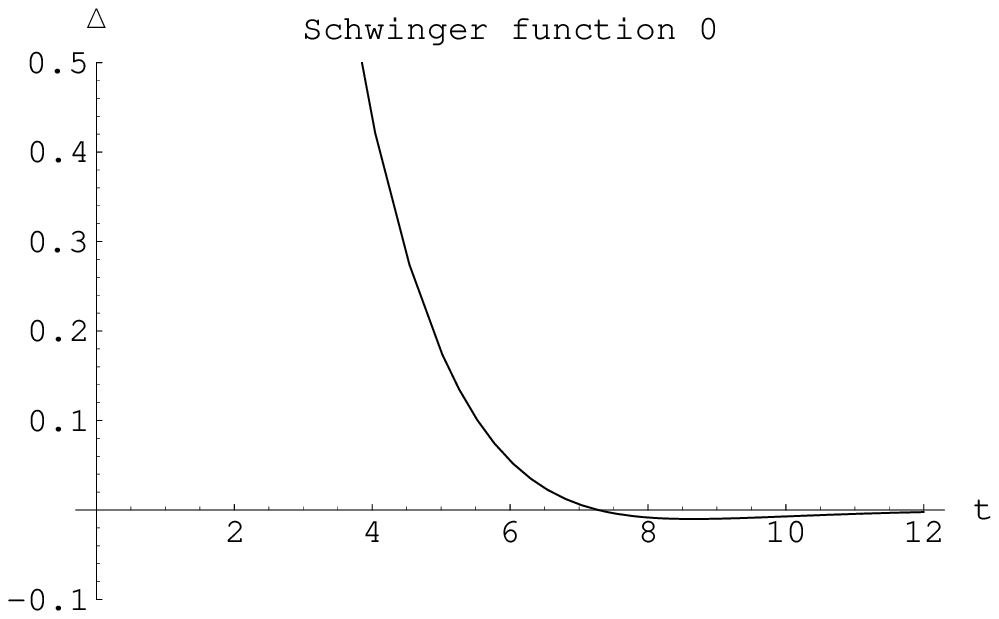}
\includegraphics[height=3.0cm,width=4.0cm]{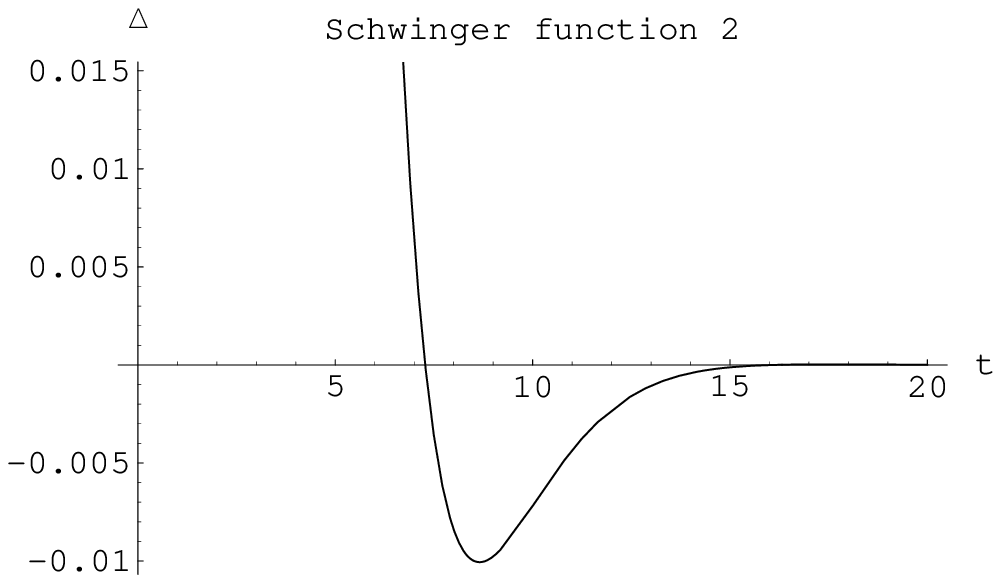}
\end{center}
\caption{
The Schwinger function $\Delta(t)$ calculated from the diagonal gluon propagator of the Gribov-Stingl type obtained in MA gauge (left panel) $0 < t< 5$, 
(middle panel) $0 < t< 12$,
(right panel) $5 < t< 15$.
}
\label{fig:Schwinger-f}
\end{figure}

\section{Summary}


In summary, we have discussed how to obtain a low-energy effective model of the $SU(2)$ Yang-Mills theory without fixing  the original gauge symmetry.
It is remarkable that the effective model respects the  $SU(2)$ gauge invariance of the original Yang-Mills theory, which allows one to take any gauge fixing in computing physical quantities of interest in the low-energy region. 
The resulting effective gluon propagator belongs to the  Gribov-Stingl type, irrespective of the gauge choice
This is a universal aspect obtained independently of the choice of gauge fixing condition.

In MA gauge, especially, we have demonstrated that the model exhibits both quark confinement and gluon confinement simultaneously in the sense that the Wilson loop average satisfies the area law (i.e., the linear quark-antiquark potential) and that the Schwinger function violates  reflection positivity.   
Moreover, we have given a formula for the string tension based on the gluon propagator of the gauge-invariant field strength $G_{\mu\nu}$.  It gives a good estimate for the string tension.

However, for the effective gluon propagator to agree exactly with the Gribov-Stingl form, we need to introduce   (i) a gauge-invariant, but nonlocal mass term or (ii) a mass term that breaks nilpotency of the BRST symmetry. 
Otherwise, we have $c_0=0$.
We argued that both quark and gluon confinement can be obtained even in the absence of such a mass term, $c_0=0$.
More results and full details will be given in a subsequent paper \cite{Kondo12}.



\section*{Acknowledgements} 
The author wishes to thank the organizers for the invitation to the very pleasant meeting in such a wonderful city.
He would also like to thank all the participants
of the workshop   for interesting discussions.

\end{document}